\documentclass{article}
\usepackage{authblk}
\usepackage[utf8]{inputenc}
\usepackage{url}
\usepackage[version=4]{mhchem}
\usepackage{multirow}
\usepackage{lipsum}
\usepackage{stackrel}
\usepackage{listings}
\usepackage{amssymb}
\usepackage{nccmath}
\usepackage{a4wide}
\usepackage{mathtools}
\usepackage{bbm}
\usepackage{amsmath,amsfonts,amsthm,amssymb}
\usepackage{bbm, dsfont}
\usepackage[justification=justified, singlelinecheck=false]{caption}
\usepackage{graphicx}
\usepackage{enumitem}
\usepackage{float}

\usepackage[justification=centering]{caption}
\usepackage{algorithmicx}
\usepackage{enumitem}
\usepackage{empheq,etoolbox}
\patchcmd{\subequations}% <cmd>
  {\theparentequation\alph{equation}}% <search>
  {\theparentequation.\alph{equation}}% <replace>
  {}{}% <success><failure>
\usepackage[justification=centering]{caption}
\usepackage{algorithmicx}
\usepackage{algorithm}
\usepackage{algpseudocode}
\usepackage{stackengine}

\usepackage[title]{appendix}
\usepackage[scaled=.90]{helvet}
\usepackage{courier}
\usepackage{ae}
\usepackage[T1]{fontenc}
\usepackage{graphicx}
\usepackage{subcaption}
\usepackage[export]{adjustbox}
\usepackage{wrapfig}

\usepackage[T1]{fontenc}
\usepackage{here} 
\usepackage[colorlinks=false, pdfborder={0 0 0}]{hyperref}

\usepackage{mdwlist}
\usepackage{tcolorbox}
\usepackage{fancybox}
\usepackage{framed}

\definecolor{darkblue}{rgb}{0,0,0.8}
\definecolor{darkgreen}{rgb}{0,0.8,0}

\definecolor{magenta}{rgb}{0.5,0,0.5}

\newcommand{\mathleft}{\@fleqntrue\@mathmargin0pt}

\renewcommand{\thefootnote}{\fnsymbol{footnote}}

\newtheorem{remark}{Remark}[section]

\usepackage{svg}
\providecommand{\keywords}[1]
{
  \small	
  \textbf{\textit{Keywords---}} #1
}

\usepackage{natbib}
\begin{document}
%\title{Exact simulation of the first-passage time of jump diffusions to time-dependent thresholds} 
\title{The hybrid exact scheme for the simulation of first-passage times of jump-diffusions with time-dependent thresholds}
\author[1]{Sascha Desmettre\thanks{Email: sascha.desmettre@jku.at}}
\author[2]{Devika Khurana\thanks{devika.khurana@jku.at}}
\author[3]{Amira Meddah\thanks{amira.meddah@jku.at}}
\affil[1]{\centerline{\small Institute of Financial Mathematics and Applied Number Theory, Johannes Kepler University, Linz}}
\affil[2]{\centerline{\small Institute of Numerical Mathematics, Johannes Kepler University, Linz}}
\affil[3]{\centerline{\small Institute of Stochastics, Johannes Kepler University, Linz}}
\date{\today}

\maketitle
\begin{abstract}
       The first-passage time is a key concept in stochastic modeling, representing the time at which a process first reaches a specified threshold. In this work, we consider a jump–diffusion (JD) model with a time-dependent threshold, providing a more flexible framework for describing stochastic dynamics. We are interested in the Exact simulation method developed for JD processes with constant thresholds, where the Exact method for pure diffusion is applied between jump intervals. An adaptation of this method to time-dependent thresholds has recently been proposed for a more general stochastic setting. We show that this adaptation can be applied to JD models by establishing a formal correspondence between the two frameworks. A comparative analysis is then performed between the proposed approach and the constant-threshold version in terms of algorithmic structure and computational efficiency. Finally, we show the applicability of the method by predicting neuronal spike times in a JD model driven by two independent Poisson jump mechanisms.
\end{abstract}

\section{Introduction}
\label{sec: intro} 
Stochastic processes are widely used to model dynamical systems in various fields to capture inherent randomness and uncertainty. A key quantity of interest in such systems is the first-passage time (FPT), representing the time at which a process first reaches a specified threshold. This concept arises naturally in various applications, for instance, in finance, it appears in the pricing of barrier options \cite{barone2008barrier}; in population dynamics, it characterizes the time for a population to reach a critical level \cite{goel2013stochastic}; and in oncology, it is used to model disease progresses beyond a critical stage \cite{roman2021using}.\\
\noindent In neuroscience, stochastic modeling plays a fundamental role in describing the intrinsic variability of neural activity resulting from complex biophysical mechanisms. Such models have a wide range of applications, including the characterization of firing rates, interspike interval distributions, and neural coding variability. Among these, a diffusion process is often used to describe the evolution of the membrane potential of a neuron \cite{braun2015first,ermentrout2010mathematical}. When this potential reaches a certain threshold, a spike is generated, and the information is transmitted to the next neuron or target cell. The time at which the membrane potential first reaches this threshold, i.e., the spike time, can be mathematically formulated as the FPT of the stochastic process.\\
\noindent To account for larger input contributions that cannot be adequately approximated within the diffusion framework, JD models were later introduced, \cite{lansky2008review,giraudo2002effects,giraudo1997jump}. These models incorporate discontinuous jumps in addition to the continuous diffusion component, allowing for a more realistic representation of neuronal activity influenced by discrete input events.\\
Analytical expressions for the FPT density of JD processes are generally unavailable, except in specific cases such as when the jump size follows a doubly exponential distribution \cite{kou2003first} or is non-negative with the boundary located below the initial value \cite{blake2003level}. The main challenge arises from the possibility of an overshoot when the process crosses the boundary. While the overshoot distribution is tractable for exponential jumps due to the memoryless property, it becomes difficult to handle for more general jump size distributions. When analytical results are unavailable, simulation-based approaches are often employed. One possibility is to simulate full sample paths using numerical methods such as Runge–Kutta or Euler-type schemes \cite{buckwar2011runge, higham2005numerical}, and compute the FPT as a by-product. Another approach avoids modeling jumps explicitly by considering a piecewise-defined threshold; in \cite{abundo2010first}, the author shows that the hitting time in this setting satisfies a certain integral equation, which is then solved numerically. A different line of work, introduced in \cite{atiya2005efficient}, proposes two Monte Carlo methods for Brownian motion. A key distinction of this method is that it requires only a few simulation points per iteration and avoids discretization bias.\\
We find the exact simulation approach particularly appealing, as it does not involve time-discretization error. This method was originally developed for pure diffusion processes, cf. \cite{herrmann2019exact,khurana2024exact}. The main idea relies on Girsanov’s transformation, which ensures that the FPT of a diffusion process has the same law as the FPT of a standard Wiener process under a different probability measure. The algorithm samples candidate FPTs from the Wiener process and accepts or rejects them with a probability derived from Girsanov’s theorem. To adapt this method to the JD setting, an algorithmic structure was established in \cite{herrmann2023exact} that applies the original idea between jumps, considering a constant threshold.\\
In many practical settings, however, the boundary is time-dependent. This is particularly relevant in neuroscience, where adaptive thresholds provide more realistic firing dynamics. To address such cases, in \cite{desmettre2025first}, the exact simulation framework was extended to piecewise diffusion Markov processes (PDifMPs) with time-dependent boundaries.\\
In this work, we establish a correspondence between PDifMPs and JD processes, allowing the exact simulation scheme developed in \cite{desmettre2025first} to be applied in the JD setting.\\
\noindent The structure of the paper is as follows. Section~\ref{setting} establishes the correspondence between the JD model and the PDifMP formulation. Section~\ref{sec: FPT of jump diffusion} presents the Exact simulation method applied from \cite{desmettre2025first} to our setting. Section~\ref{sec:Comp.} analyzes the computational properties of the proposed approach and compares it with the constant-threshold case. Finally, Section~\ref{sec:application} illustrates the applicability of the method through a stochastic neuron model with two independent sources of jumps. 

\section{Model structure and reformulation via PDifMPs}
\label{setting}
In this section, we build upon the work presented in \cite{herrmann2023exact} and reformulate their JD model within the PDifMP framework. While both settings feature a combination of continuous dynamics and discontinuous jumps, a PDifMP provides a more structured and hybrid representation, in which the continuous and discrete components evolve on well-separated time scales and are governed by distinct rules. In particular, the continuous dynamics evolve according to a stochastic differential equation (SDE) between random jump times, while the discrete changes are governed by a state-dependent jump rate and a transition kernel. This separation of dynamics allows the model to be interpreted as a hybrid Markov process with piecewise diffusion trajectories interrupted by stochastic resets. For a broader theoretical foundation of such hybrid representations and their applications, we refer the reader to \cite{meddah2024stochastic, bect2007processus, buckwar2023stochastic, buckwar2025american, buckwar2025numerical}. This approach is particularly advantageous because it allows us to treat the continuous dynamics and the jump mechanisms on separate, independent intervals, thereby reducing the complexity of finding the FPT for the JD SDE and connecting it to the framework introduced in \cite{khurana2024exact, desmettre2025first}.\\
We start by recalling the JD model introduced in \cite{herrmann2023exact}. Let $\big(\Omega, \mathcal{F}, (\mathcal{F}_t)_{t\geq 0}, \mathbb{P}\big)$ be a filtered probability
space satisfying the usual conditions and let $\mathcal{E} \subset \mathbb{R}\setminus\{0\}$ denotes the mark space, that is, the set that defines the possible locations or outcomes of the jumps, \cite{bruti2007strong}. Consider a Poisson random measure $p_{\phi}(d\eta \times dt)$ on $\mathcal{E} \times [0, \infty)$ with intensity measure $\phi(d\eta) dt$, where $\phi$ is a finite, non-negative measure. The associated total jump intensity is $\lambda = \phi(\mathcal{E})$.\\
The dynamics of a JD process $Z_t$ are governed by the following jump SDE
\begin{equation}
    dZ_{t} = \mu(t,Z_{t-})\,dt + \sigma(t,Z_{t-})\,dB_t + \int_{\mathcal{E}} j\big(t,\,Z_{t-},\,\eta \big)\, p_{\phi}(d\eta \times dt), \quad Z_0 = z_0 \in \mathbb{R}.
    \label{eq:jump_diffusion_sde}
\end{equation}
Here, $B_t$ is a standard Brownian motion and $Z_{t^{-}}$ refers to the value of $Z_{t}$ just before time $t$. The function $\mu(t,\,Z_{t^{-}})$ represents the drift of the process and $\sigma(t,\,Z_{t-})$ is the diffusion coefficient. The integral term represents the jump contribution, where $j(t,\, Z_{t^-},\, \eta)$ denotes the jump size due to a Poisson event with characteristic $\eta \in \mathcal{E}$ at time $t$. 
\begin{remark}
The Poisson random measure $p_\phi$ implicitly encodes the entire jump structure of the process through the following components:
\begin{enumerate}
    \item the jump counting process $N_t := p_\phi(\varepsilon \times (0,t])$;
    \item the jump magnitudes, represented by the map
    \begin{equation}
    \eta \mapsto j(T_i, Z_{T_i^-}, \eta_i),\qquad \eta_i \in \mathcal{E}, 
    \label{jump_mark}
    \end{equation}
    where $(T_i,\eta_i)$ denotes the $i$-th marked jump event.\\
\end{enumerate}
\end{remark}
\noindent While this measure-theoretic formulation is compact, it hides the discrete-event structure of the jumps. 
In particular, extracting explicit information about jump times and magnitudes, which is essential, for instance, when studying FPT, requires decomposing the Poisson measure into a sequence of $(T_i,z_i)$. 
This involves identifying both the random jump times $T_i$ and their associated marks $z_i$, and then applying the state-dependent mapping \eqref{jump_mark}.\\
\noindent To explicitly characterize the discrete–continuous interaction and facilitate simulation of jump times, 
we reformulate the system \eqref{eq:jump_diffusion_sde} as a PDifMP. The full state variable is represented by the following couple of processes:
\begin{equation}
    Z_t = (Y_t,\, N_t) \in E := \mathbb{R} \times \mathbb{N},
    \label{PDif_pro}
\end{equation}
here $Y_t$ denotes the continuous component, and the discrete part is given by the homogeneous Poisson process $(N_t)_{t\ge0}$ with rate $\lambda>0$, independent of the Brownian motion $(B_t)_{t\ge0}$.\\
\noindent We denote the jump times by $0=:T_0<T_1<T_2<\dots$, with i.i.d. interarrival durations $  e_i := T_i - T_{i-1} \sim \mathrm{Exp}(\lambda)$.\\
\noindent Let $(\zeta_i)_{i\ge1}$ be i.i.d. random marks with law $\nu$ on $\mathcal{E}$, independent of $(B_t,N_t)$. On each interval $[T_i,T_{i+1})$, the continuous component evolves according to
\begin{equation*}
dY_t^{i} = \mu(t,Y_t^{i})\,dt + \sigma(t,Y_t^{i})\,dB_t, 
\qquad Y^{i}_{T_i} = y_{\tiny{T_i}}, \quad t \in [T_i,T_{i+1}),
\end{equation*}
while $N_t \equiv i$ remains constant on $[T_i,T_{i+1})$.\\
\noindent At a jump time $T_i$, the state updates as
\begin{equation*}
Y_{T_i} = Y_{T_i-} + j(T_i,Y_{T_i-},\zeta_i), 
\qquad 
N_{T_i} = N_{T_i-} + 1,
\end{equation*}
where $j:\mathbb{R}_+\times\mathbb{R}\times\mathcal{E}\to\mathbb{R}$ is the jump-size map.  We assume $U_0=(Y_0,0)$ with $Y_0$ given.\\
The post-jump distribution is described by a Markov transition kernel $K:\mathbb{R}\times\mathcal{B}(\mathbb{R})\to[0,1]$ given by
\begin{equation}
   K\big((y,n),\, A\times\{n+1\}\big)
= \int_{\mathcal{E}} \mathbf{1}_{A}\!\,\Big(y+j(t,y,\zeta)\Big)\,\nu(d\zeta),
\qquad 
K\Big((y,n),\, \mathbb{R}\times\{n\}\Big)=0. 
\label{transition_kernel}
\end{equation}
If the jump map is independent of $t$ and $j(y,\zeta)=\zeta$ (purely additive jumps), then for any $A\in\mathcal{B}(\mathbb{R})$ the post-jump distribution of the continuous state reduces to
\begin{equation*}
    \mathcal{Q}(y,A)
    := \mathbb{P}\big(Y_{T_i}\in A \,\big|\, Y_{T_i-}=y\big)
    = \nu(A-y),
\end{equation*}
which is the distribution of $y+\zeta$ with $\zeta\sim\nu$. \\
\noindent This reformulation establishes a rigorous correspondence between the JD model \eqref{eq:jump_diffusion_sde} and the PDifMP representation \eqref{PDif_pro}. 
The main structural advantage is the explicit separation of continuous and discrete dynamics: the continuous component evolves as a diffusion between successive jump times, while the discrete component records the jump events. \\
This decoupling facilitates the use of existing methods for the analysis of FPTs, such as those introduced in \cite{khurana2024exact}. 
Moreover, it provides a modular framework that not only simplifies exact simulation of the first-passage time (see Section~\ref{section: exact app}), but also allows for extensions to more general jump mechanisms, while retaining analytical tractability.

\section{The FPT of jump diffusions}
\label{sec: FPT of jump diffusion}
In this section we define the FPT of the JD \eqref{eq:jump_diffusion_sde} with respect to a time-dependent threshold. We then present a simulation approach, developed for PDifMPs in \cite{desmettre2025first}, which applies directly in our setting via the correspondence established in Section~\ref{setting} between the JD and PDifMPs. The underlying algorithmic structure builds on the idea of employing the exact simulation method for pure diffusions, originally proposed in \cite{herrmann2023exact} for constant thresholds.\\
\noindent  Let $\theta:\mathbb{R}^+_0\to\mathbb{R}$ be a deterministic, continuous threshold function. We define the FPT of the JD \eqref{eq:jump_diffusion_sde} to the boundary $\theta$ as the random variable
\begin{equation}
    \tau_\theta := \inf\{t > 0 : Z_t = \theta(t)\}.
    \label{eq:fpt}
\end{equation}
\noindent The explicit distribution of $\tau_\theta$ is in general analytically intractable, due to the interplay between the continuous stochastic fluctuations, the discontinuities induced by jumps, and the time dependence of the boundary $\theta$.\\
In the following subsection, we present the method based on Exact Simulation approach that can be applied in our setting to generate approximated samples of FPT of $\tau_\theta$. The significance of this method lies in the fact that it does not involve time-discretization error.
\subsection{Exact simulation approach}
\label{section: exact app}

\setcounter{footnote}{0}
\renewcommand{\thefootnote}{\alph{footnote}}

In this section, we describe the hybrid exact FPT simulation scheme (HEx Scheme) and apply it to the jump–diffusion model \eqref{eq:jump_diffusion_sde} to compute $\tau_\theta$. We first recall the algorithmic structure introduced in \cite{herrmann2023exact}, which extends the Exact Simulation method for pure diffusions (cf.~\cite{herrmann2019exact}) to systems with jumps. We then outline how each stochastic component of the algorithm is simulated, following the framework proposed in \cite{desmettre2025first}.\\
\noindent The HEx Scheme applies the Exact simulation method between successive jump times to detect the first hitting to the threshold. The underlying Exact method relies on a change of measure via Girsanov’s transformation, under which the diffusion process is represented in terms of a standard Brownian motion serving as an auxiliary process. This allows the simulation of diffusion paths without time discretization by sampling Brownian trajectories and accepting or rejecting them according to a likelihood ratio derived from Girsanov’s theorem (see, e.g., \cite{jeanblanc2009mathematical}). Since this version of Girsanov’s theorem is valid only for continuous diffusion processes, it is applied within each inter-jump interval, where the dynamics are purely continuous.\footnote{Please refer to \cite{DesmettreLeobacherRogers2021ChangeOfDrift} for the change of measure in a general one-dimensional diffusion setting.}\\
 To do this, we define the \textit{tracking process} $Y^{i,\infty}$ from the continuous part of the process \eqref{eq:jump_diffusion_sde}  as the solution to the following SDE
    \begin{equation}
    dY^{i,\infty}_t = \mu\big(t,\, Y^{i,\infty}_t\big)\, dt + \sigma\big(t,\,Y^{i,\infty}_t\big)\, dB_t, \qquad t \in [T_i, \infty), \, \, T_0=0,
    \label{eq:tracking process}
    \end{equation}
with $Y^{i,\infty}_{T_i} = Y_{T_i}:=y_i$. Defining this tracking process, enables us to use the Exact method in the HEx Scheme. If the FPT of the tracking process $Y^{i,\infty}$ occurs before JD gets a jump, then $\tau_\theta$ should be equivalent to FPT of $Y^{i,\infty}$.\\
Let us define FPT of the process $Y^{i,\infty}$ as 
\begin{equation*}
        \tau_{b}^{\text{cont}, i} := \inf\left\{ t > T_i \, \middle| \, Y^{i,\infty}_t = \theta(t) \right\}.
\end{equation*}
Under mild conditions on $\sigma$, in particular that $\sigma$ is non-negative and integrable, the Lamperti transform, cf. ~\cite{panik2017stochastic} can be applied as follows:
\begin{equation*}
  X^{i,\infty}_t:=F^{i}\big(t,\, Y^{i,\infty}_{t}\big) := \int^{y} \frac{1}{\sigma\big(t, \, h\big)} dh \bigg|_{y=Y^{i,\infty}_t},
\end{equation*}
which transforms the SDE \eqref{eq:tracking process} to
\begin{equation}
dX^{i,\infty}_t = \alpha\big(t,\, X^{i,\infty}_t\big)dt + dB_t, \quad   X^{i,\infty}_{T_{i}} = F^i\big(T_i,\, Y^{i,\infty}_{T_{i}}\big),
\label{eq:cont_SDE_after_Lamperti}
\end{equation}
where 
\begin{equation*}
    \alpha\big(t,\, x\big) = \left.\left(\frac{\partial F^i}{\partial t}\big(t,\, y\big)+ \frac{\mu\big(t,\, y\big)}{\sigma\big(t,\, y\big)} - \frac{1}{2}\frac{\partial \sigma}{\partial y}\big(t,\, y\big) \right)\right|_{y={(F^i)}^{-1}(t,\, x)}.
\end{equation*}
\begin{remark}
\begin{enumerate}
    \item Since the Lamperti transform is bijective in the continuous variable $y$, computing the FPT of $Y^{i,\infty}$ to the threshold $\theta(t)$ is equivalent to computing the FPT of $X^{i,\infty}$ to the transformed threshold
    \begin{equation*}
        \beta(t) := F^i\big(t,\, \theta(t)\big).
    \end{equation*}
    \item We use the notation $\tau_{\beta}^{\text{cont},i}$ for the FPT of the process $X^{i,\infty}$ to $\beta$.
    \item Girsanov's transformation facilitates a change of measure under which the transformed process \eqref{eq:cont_SDE_after_Lamperti} has the same law as a Brownian motion under a different measure.
\end{enumerate}
\end{remark}
\begin{algorithm}[ht]
\caption{Schematical representation of the HEx Scheme}
\label{alg:HEx Scheme}
\begin{algorithmic}[1]
    \State Initialize: set $i=0$, $T_0=0$, $Y_0=y_0$, $\tau_b=0$, and ensure $X_0^{0,\infty}<\beta(0)$.
    \Repeat 
        \State Simulate $\tau_\beta^{\text{cont},i}$, the FPT of the continuous SDE \eqref{eq:tracking process}, using the Exact algorithm in \cite{khurana2024exact}.
        \State Simulate the next jump time $T_{i+1}=T_i+e_i$, where $e_i\sim\mathrm{Exp}(\lambda)$.
        \If{$\tau_\beta^{\text{cont},i}<T_{i+1}$}  
            \State Set $\tau_b=\tau_\beta^{\text{cont},i}$ and \textbf{stop}.
        \Else 
            \State Simulate $X^{i,\infty}_{T_{i+1}}$ and apply the inverse Lamperti transform:
            \begin{equation*}
                Y^{i,\infty}_{T_{i+1}} = {(F^i)}^{-1}\!\big(T_{i+1},X^{i,\infty}_{T_{i+1}}\big).
            \end{equation*}

            \State Sample a jump component $\zeta_i\sim\nu$ and update the process as:
            \begin{equation*}
                Y_{T_{i+1}} = Y^{i,\infty}_{T_{i+1}} + j(T_{i+1},Y^{i,\infty}_{T_{i+1}},\zeta_i).
            \end{equation*}

            \State Apply the Lamperti transform again to initialize the next iteration:
            \begin{equation}
                 X^{i+1,\infty}_{T_{i+1}} = F^{i+1}(T_{i+1}, Y_{T_{i+1}}).
            \end{equation}
            \If{$X^{i+1,\infty}_{T_{i+1}} \geq \beta(T_{i+1})$} 
                \State Set $\tau_b=T_{i+1}$ (threshold crossed due to the jump) and \textbf{stop}.
            \Else 
                \State Continue to the next iteration with $i \leftarrow i+1$.
            \EndIf
        \EndIf
    \Until $\tau_b \neq 0$.
\end{algorithmic}
\end{algorithm}
The core idea of the HEx Scheme, outlined in Algorithm \ref{alg:HEx Scheme}, is to iteratively check for threshold crossings, both within each inter-jump interval and at the jump times themselves.\\ 
At each step, the continuous dynamics are represented by the tracking process, whose FPT to the threshold $\beta$,
\begin{equation}
    \tau_{\beta}^{\text{cont},i}=\big\{t\in [T_{i},\infty) \mid X^{i,\infty}_t = \beta(t)\big\},
    \label{eq: first_component}
\end{equation}
can be simulated exactly using the Exact algorithm of \cite{khurana2024exact}.\\
\noindent If this hitting time occurs before the next jump, it coincides with the FPT $\tau_b$ of the JD. 
Otherwise, the process is advanced to the jump time $T_{i+1}$, the jump is applied, and the updated state is used to initialize a new tracking process.\\
This requires evaluating the transformed process $X^{i,\infty}$ at $T_{i+1}$ under the condition $\tau_\beta^{\mathrm{cont},i}\geq T_{i+1}$, 
\begin{equation}
    \Big\{X^{i,\infty}_t \mid \tau_\beta^{\mathrm{cont},i}\geq T_{i+1}\Big\}_{t\in[T_i,\infty)} \quad \text{at } T_{i+1}, 
    \label{eq: second component}
\end{equation}
a step carried out by the exact simulation procedure of \cite{desmettre2025first}, summarized in Algorithm~\ref{alg:Y_at_jump}. 
The procedure then continues recursively until the threshold is crossed.
\begin{algorithm}[ht]
    \caption{Schematical description of the Exact simulation of $X^{i,\infty}_{T_{i+1}}$ given $\tau^{\mathrm{cont},i}_{\beta} \geq T_{i+1}$}
    \label{alg:Y_at_jump}
    \begin{algorithmic}[1]
        \State Initialize parameters: minimal slope $s_\text{min}$, tolerance $\epsilon$.
        \State \textbf{Step 1:} \textit{Simulate a point from the conditional Brownian path.}
            \State \quad \textbf{Step 1.1} Construct a tilted line with slope $s > s_\text{min}$ below the threshold $\beta$ with initial value $\beta(T_i)$. Simulate the first-passage time \(t_1\) of the standard Wiener process \(W\) to this line using its explicit density.
            \State \quad \textbf{Step 1.2}     \textbf{while} $t_j < T_{i+1}$ \textbf{do}
                 \State \quad \quad \textbf{if} {$t_1 > T_{i+1}$ and slope $s > s_\text{min}$}  \textbf{then}  
                 \State \quad \quad \quad Reduce the slope $s$ and restart simulation from $t_1$.
                 \State \quad \quad \textbf{end if}
                 \State \quad \quad \textbf{if} distance $|\beta(t_j) - W_{t_j}| < \epsilon$ \textbf{then} \State \quad \quad \quad Restart simulation from $t_1$.
                 \State \quad \quad \textbf{end if}
                 \State \quad \quad Construct a new tilted line with the same slope $s$ starting at $(t_j, \beta(t_j))$.
                 \State \quad \quad Simulate the next first-passage time $t_{j+1}$ to this new line. 
        \State Output of Step 1 : a point $(t_c,\beta_c(t_c))$ such that $t_c \geq T_{i+1}$ and $\beta_c$ is the constructed tilted line that the Brownian motion hits at $t_c$ for the first time.        
        \State \textbf{Step 2:} \textit{Construct the auxiliary Brownian path.}  
        We found a point from the auxiliary path in the previous step, we construct the full auxiliary Brownian path up to time $t_c$ using a Bessel bridge.
        \State \textbf{Step 3:} \textit{Rejection sampling.}  
        Apply the rejection probability from \cite{beskos2005exact} to accept or reject the constructed path.
        \State \quad \textbf{Step 3.1:} \textit{Extract $X^{i,\infty}_{T_{i+1}}$ using thinning.}  
        Apply thinning split over the intervals $[T_i, T_{i+1}]$ and $[T_{i+1}, t_c]$, to obtain the value $X^{i,\infty}_{T_{i+1}}$ from an accepted path.
    \end{algorithmic}
\end{algorithm}
The algorithm builds on the classical Exact method while specifically addressing the additional challenge posed by the condition $\tau_b^{\text{cont},i} \geq T_{i+1}$. Like the classical method, it employs rejection sampling, i.e., a Brownian path is treated as an auxiliary process, and the corresponding rejection probability is evaluated to accept or reject the sample path. Because of the condition, the auxiliary process should be a conditional Brownian motion.\\
\noindent In Step 1 of the Algorithm \ref{alg:Y_at_jump}, we are simulating a point from a sample path of auxiliary process (conditional Brownian motion). We achieve this by constructing successive tilted lines below the threshold and simulating FPTs, $t_j$, of the auxiliary process to these lines. Each new tilted line is constructed starting from $\beta(t_j)$. To ensure that Brownian motion satisfies the condition, we check the distance between the Brownian motion and the threshold $\beta$ at each FPT. If the distance is too small, i.e., less than a small parameter $\epsilon$ before the jump time, then this indicates that the Brownian path has reached the threshold before the jump time. In that case, we start with a new Brownian path. We simulate FPTs of these constructed lines until we find one which is greater than or equal to $T_{i+1}$.

\noindent Now that a valid point from the auxiliary path has been sampled, the next step is to simulate the entire Brownian trajectory on the time interval $[T_i, t_c]$ that stays below the threshold $\beta$ and passes exactly through this point $(t_c, \beta_c(t_c))$.

\noindent To construct such a path, we use the fact that the process
\begin{equation*}
  R_h = \beta(t_c - h) - W_{t_c - h}, \qquad 0 \leq h \leq t_c - T_i,  
\end{equation*}
is a Bessel bridge, \cite{hernandez2013hitting}. We begin by simulating a Bessel bridge $R_h$ over $[0, t_c - T_i]$ with initial value
\begin{equation*}
R_0 = \beta(t_c) - \beta_c(t_c),    
\end{equation*}
and terminal value
\begin{equation*}
 R_{t_c - T_i} = \beta(T_i) - Y^{i,\infty}_{T_i}.   
\end{equation*}
Once the Bessel bridge $R_h$ is generated, using its definition that it is a norm of 3 dimensional Brownian motion, cf. \cite{oksendal2013stochastic,revuz2013continuous}, we invert the transformation to recover the conditional Brownian path $W_t$ over $[T_i,t_c]$. This path serves as the auxiliary process for the rejection sampling procedure. Note that this resulting process would be a conditional Brownian motion since it satisfies the condition that its hitting time occurs after the jump time $T_{i+1}$. Let us denote this conditional Brownian motion as $W^i$, where $i$ indicates that it initializes with $X_{T_{i}}$ at $T_i$.

\noindent In the last step, we use the rejection probability:
\begin{equation}
\phi(t) = \exp\left(
A\left(t_c,\,W^i_{t_c}\right)
- \int_{T_i}^{t_c} \gamma\left(t,\, W^i_t\right) \, dt
\right)
\end{equation}
from \cite{beskos2006retrospective} to perform rejection sampling, where
\begin{align}
A\big(t, x\big) &:= \int^x \alpha\big(t, \, h\big)\, dh, \notag \\[0.22cm]
\gamma\big(t,\, x\big) &:= \frac{\partial}{\partial t} A\big(t,\, x\big) + \frac{1}{2} \left( \frac{\partial \alpha}{\partial x}\big(t,\, x\big) + \alpha^2\big(t,\, x\big) \right).
\label{eq: gamma}
\end{align}
Directly evaluating the probability $\phi(t)$ is not feasible in closed form, as it depends on entire continuous trajectory of $W^i$.
To address this, we apply a Poisson thinning technique, see for e.g. \cite{beskos2006retrospective} that only requires evaluation at finitely many time points. Since we are interested in the value of the process at $T_{i+1}$, we split the probability over two intervals $[T_i,T_{i+1}]$ and $[T_{i+1},t_c]$. This is because, when we use thinning, it is not guaranteed that those finitely many sampled points also include evaluation at the time point $T_{i+1}$.
\begin{remark}
\begin{enumerate}
    \item In Step~1.2 of Algorithm~\ref{alg:Y_at_jump}, the slope $s$ of the tilted line must be chosen carefully. 
If the simulated hitting time $t_1$ to the first line starting at $\beta(T_i)$ exceeds $T_{i+1}$, the slope is reduced and a new $t_1$ is generated. 
This procedure may be repeated several times, so a lower bound $s_{\min}$ is imposed to prevent infinite reductions. 
\item Fixing a single slope from the start would correspond to a fixed linear threshold, since the intercept for the first constructed threshold is always taken as $\beta(T_i)$. Setting $t_c \gets t_1$ then implicitly assumes the hitting time to this linear threshold is finite. To avoid such bias, the slope of the tilted lines is chosen depending on whether $t_1\geq T_{i+1}$.
\item  The HEx Scheme coincides with the method developed for PDifMPs in \cite{desmettre2025first}. In particular, the JD model \eqref{eq:jump_diffusion_sde} represents a special case of a PDifMP where the jump times follow a Poisson process, the transition kernel is given by \eqref{transition_kernel}, and the drift and diffusion coefficients depend only on the continuous component of the process, rather than explicitly on the jump component.
\end{enumerate}
\end{remark}

\section{Comparison analysis with Herrmann and Massin's approach}
\label{sec:Comp.}
The structure of the HEx Scheme is closely related to the exact simulation framework of \cite{herrmann2023exact}.
In what follows, we clarify the key algorithmic differences, their consequences for bias and computational complexity, and provide a numerical example for comparison. 
\subsection{Algorithmic differences}
\noindent The differences in the algorithm arise from the way the Exact simulation method is adapted to simulate the two random components in \eqref{eq: first_component} and \eqref{eq: second component}. These modifications are necessary to handle a more general, time-dependent threshold.
\begin{itemize}
    \item[(i)] \emph{Modification in sampling $\tau_{\beta}^{\text{cont},i}$}: The explicit density of the auxiliary process, a standard Brownian motion with initial value $X^{i,\infty}_{T_i}$ at $T_i$, is available for constant, linear, or piecewise linear thresholds. For more general thresholds, we employ an efficient approximation method from \cite{herrmann2016first}.\\
    Additionally, the rejection probability is updated to account for the time-dependent nature of the threshold, cf. \cite{khurana2024exact}.
    \item[(ii)] \emph{Modification in sampling $X^{i,\infty}_{T_{i+1}}$ from \eqref{eq: second component}}: The density of the auxiliary process, the conditional Brownian motion
    \begin{equation}
    W^i=\{W_t \mid \tau_\beta^W\geq T_{i+1}\}, \quad W^i_{T_i}=X^{i,\infty}_{T_i},
    \label{eq: conditional brownian motion}
    \end{equation}
    is available explicitly only for a constant threshold. Here, $W$ is a standard Brownian motion and $\tau_\beta^W$ is the FPT of $W^i$ corresponding to $\beta$. In our setting, this gives rise to Step~1 of Algorithm~\ref{alg:Y_at_jump}, where successive tilted lines are generated until a valid post-jump contact point with the process \eqref{eq: conditional brownian motion} is identified.
    The rejection probability remains the same, but rejection sampling is performed up to the contact time point $t_c$, chosen from Step~1, since the whole dynamics of the auxiliary process is constructed using the point $(t_c, \beta_c(t_c))$. In contrast, the method in \cite{herrmann2023exact} only requires evaluation up to time $T_{i+1}$.
\end{itemize}

\subsection{Consequences for bias and complexity} 
The algorithm of \cite{herrmann2023exact} introduces no approximation error beyond the inherent statistical error for constant thresholds. In our case, two parameters introduce bias:
\begin{itemize}
    \item $s_{\min}$, the minimal slope used in constructing tilted thresholds in Step~1 of Algorithm \ref{alg:Y_at_jump}. If no restriction was placed on the minimum slope, i.e., if $s_{\text{min}}=-\infty$, the constructed tilted line could take any slope, and the algorithm would, in principle, be free of this source of bias.
    \item $\epsilon$, the tolerance parameter to detect threshold crossing. As shown in \cite{herrmann2016first}, decreasing $\epsilon$ ensures convergence, and its effect on runtime is mild.
\end{itemize}
The influence of $s_{\min}$ was studied in \cite{desmettre2025first}, where both a convergence proof and numerical validation were provided. \\

\noindent In the following, we address two main points. 
First, we provide a theoretical comparison of the computational effort required by our algorithm versus the method of \cite{herrmann2023exact}. 
In particular, we analyse how many additional iterations are introduced when adapting the algorithm to handle a time-dependent threshold. \\
Second, we consider a numerical example with a constant threshold. In this setting, we apply HEx Scheme to simulate FPTs, which naturally introduces bias due to the parameters $s_{\text{min}}$ and $\epsilon$. We then compare these results with the FPTs obtained using the exact method of \cite{herrmann2023exact}. This comparison validates the accuracy of the scheme and also helps to find out if it is difficult to detect the suitable values of $s_\text{min}$ and~$\epsilon$. 
\paragraph{\textit{Iteration complexity}}
In the previous subsection, we discussed the additional steps introduced by the HEx Scheme, which naturally lead to an increased time complexity:
\begin{itemize}
    \item[(i)] sampling $\tau_\beta^W$ for rejection sampling to compute $\tau_{\beta}^{\text{cont},i}$, this is discussed in \cite{khurana2024exact},
    \item[(ii)] Step~1 of Algorithm \ref{alg:Y_at_jump} for simulating a point from a sample path of the auxiliary process $W^i$, and
    \item[(iii)] rejection sampling in Step~3 of Algorithm \ref{alg:Y_at_jump}, which in our case is performed on the time interval $[T_i, t_c]$ instead of $[T_i, T_{i+1}]$ as in \cite{herrmann2023exact}.
\end{itemize}
Note that we refer here to the additional time complexity associated with a single jump interval $(T_{i},T_{i+1}]$.\\
We first derive a bound on the expected number of iterations required in Step~1 of the Algorithm \ref{alg:Y_at_jump} to construct a valid tilted line that yields a contact point $(t_c,\beta_c(t_c))$ with $t_c \geq T_{i+1}$.\\
Our tilted-line construction is inspired by the approach in \cite{herrmann2016first}, with two additional corrections 
\begin{itemize}
    \item[(i)] the Brownian motion $W^i$ must satisfy the condition $\tau_\beta^W\geq T_{i+1}$; and
    \item[(ii)] the slope $s$ is adapted if $t_1^{(s)} \geq T_{i+1}$, where $t_1^{(s)}$ denotes the hitting time of the Brownian motion to the first constructed linear threshold with slope $s$. 
\end{itemize}

\noindent Let, $\{s_k\}$ be a decreasing sequence of slopes to consider such that $s_k>s_{k+1}$ until $s_{\text{min}}$ is reached. Note that there are different possibilities of choosing the sequence and that does not impact the correctness of the algorithm.\\
We start by simulating $t_1^{(s_1)}$, and the expected number of iterations at this step is given by
\begin{equation*}
\mathbb{P}(t_1^{(s_1)} < T_{i+1})\cdot \underset{\text{no more iteration}}{\underbrace{0}} + \mathbb{P}(t_1^{(s_1)}\geq T_{i+1})\cdot \underset{\text{+1 iteration to reduce slope}}{\underbrace{1}}.    
\end{equation*}

This procedure continues until we find a slope small enough such that $t_{1}^{(s_k)}<T_{i+1}$. Hence, the expected number of iterations $I_1$ required to find such a $t_1^{(s_k)}$ is given by
\begin{equation*}
  \mathbb{E}[I_1] = \sum_{k:s_k\geq s_{\text{min}}} \mathbb{P}(t_1^{(s_k)}\geq T_{i+1}).    
\end{equation*}

Note that $t_1^{(s_k)}$ is the FPT of a Brownian motion to a linear threshold with slope $s_k$ and initial value $\beta(T_i)$; therefore, its explicit density exists (cf.~\cite{karatzas1991brownian}).\\
In \cite{herrmann2016first}, a bound was established for the expected number of iterations that is required to construct the tilted horizontal lines, until hitting time $\tau_\beta^W$ occurs. Specifically, there exists positive constants $C_1$, $C_2$, $k_1$, $k_2$ and $\epsilon_0$ such that for any $k_1\geq$ upper bound of $\beta'(t)$ and any $(T, -s_\mathrm{min})$ satisfying $(-s_\mathrm{min}+k_1)T\leq k_2$, we have
\[
   \mathbb{E}[I_2] \leq (C_1-C_2 s_\mathrm{min}) |\log\epsilon| \quad \forall \epsilon \leq \epsilon_0,
\]
where $T\leq \tau_\beta^W$ is the terminal time of the algorithm.\\
In our setting, the two procedures of choosing a suitable $t_1$ and the tilted-line construction are repeated until the condition $\tau_\beta^W\geq T_{i+1}$ is satisfied.
Thus, the total expected number of iterations for Step~1 of Algorithm \ref{alg:Y_at_jump} is
\begin{equation*}
 \mathbb{E}[I] \;\leq\; 
\left( \mathbb{E}[I_1]
\;+\; \mathbb{E}[I_2]\right)
\cdot \frac{1}{\mathbb{P}(\tau_\beta^W \geq T_{i+1})}. 
\end{equation*}
\noindent Here, the denominator represents the success probability, implying that the expected number of draws until success is $1/\mathbb{P}(\tau_\beta^W \geq T_{i+1})$. Since $\beta$ is time dependent, this probability has no closed form in general, but one may use the bound,
\begin{equation*}
 \mathbb{P}(\tau_\beta^W \geq T_{i+1}) 
\;\geq\; \mathbb{P}(\tau_{\beta_{\min}}^W \geq T_{i+1}),   
\end{equation*}

\noindent where $\beta_{\min} := \inf_{t \in [T_i,\infty]} \beta(t)$, if $\beta_\mathrm{min}$ exists. The distribution of $\tau_{\beta_{\min}}^W$ is explicitly known.  

\paragraph{\textit{Extended rejection sampling}} A further time complexity difference arises in Step~3 of Algorithm~\ref{alg:Y_at_jump}. The expected number of iterations for this step is of order
\begin{equation}
   e^{\kappa(T_{i+1}-T_{i})-A(T_{i+1}, W^i_{T_{i+1}})} \cdot e^{\kappa(t_c - T_{i+1})-A(t_c, W^i_{t_c})} 
   \label{eq: exp_itrtns_thinning}
\end{equation}
compared to $e^{\kappa (T_{i+1}-T_i)-A(T_{i+1}, W^i_{T_{i+1}})}$ in \cite{herrmann2023exact} for this step. The expected number of iterations for performing rejection sampling via the thinning procedure grows exponentially with $\kappa$ and the length of the time interval (cf.~\cite{khurana2024exact,herrmann2019exact}). In Step~3, two thinning procedures are applied over the intervals $[T_i,T_{i+1}]$ and $[T_{i+1},t_c]$, resulting in the expected iteration count given above in \eqref{eq: exp_itrtns_thinning}. Here, $\kappa$ denotes the upper bound for the function $\gamma$ defined in \eqref{eq: gamma}, i.e.
\begin{equation}
    \kappa \geq \sup_{(t,x)\in \mathbb{R}^+_0\times \mathbb{R}} \gamma(t,x)
    \label{def:kappa}
\end{equation}
\subsection{Numerical comparison}

Consider the following JD:
\begin{equation}
        dZ_{t} = \big(1.6 + \sin(Z_t)\big)\,dt + dB_t 
    + \int_{\mathcal{E}} j(t,Z_{t-}, \eta)\, p_{\phi}(d\eta \times dt), 
    \qquad Z_0 = -1,
    \label{JD_exp1}
\end{equation}

with jump function $j(t,z,x) = z - x\sin(z)$ and constant threshold $\theta \equiv 1$.\\
In this subsection, we compare our modified approach with that of Herrmann and Massin, which serves as a reference method. Since the approach of Herrmann and Massin introduces no approximation error, it provides reliable means to validate the correctness of our modified algorithm described in Algorithm~\ref{alg:Y_at_jump}.\\
\noindent We first start by writing the JD model \eqref{JD_exp1} as a PDifMP. Let $(\Omega,\mathcal{F},(\mathcal{F}_t)_{t\ge0},\mathbb{P})$ be a filtered probability space and let $(B_t)_{t\ge0}$ be a one–dimensional Brownian motion and $N=(N_t)_{t\ge0}$ a Poisson process of rate $\lambda$, independent of $B_t$. The PDifMP state is $Z_t=(Y_t,N_t)\in E:=\mathbb{R}\times\mathbb{N}$, with $Y_0=-1$.\\
\noindent On each inter–jump interval $[T_i,T_{i+1})$,
the continuous component $Y$ solves the diffusion SDE
\begin{equation*}
dY_t \;=\; \mu(Y_t)\,dt + dB_t,
\qquad \mu(y):=1.6+\sin y,
\qquad t\in[T_i,T_{i+1}),    
\end{equation*}

with initial condition $Y_{T_i}$ at time $T_i$, $T_0=0$ by convention.\\
The jump times $(T_i)_{i\ge1}$ are the arrival times of $N_t$.
Hence the jump rate function is given by $\lambda(z)\equiv \lambda, z\in E$.\\
At a jump time $T_i$ draw an i.i.d. mark $\eta_i\sim\nu$ (independent of the past), and update as follows
\begin{equation*}
Y_{T_i}
\;=\;
Y_{T_i-} + j(T_i,Y_{T_i-},\eta_i),
\qquad
j(t,y,\eta) := y - \eta\,\sin y,
\qquad
N_{T_i}=N_{T_i-}+1.    
\end{equation*}

\begin{figure}[H]
    \centering
    \includegraphics[width=0.7\linewidth]{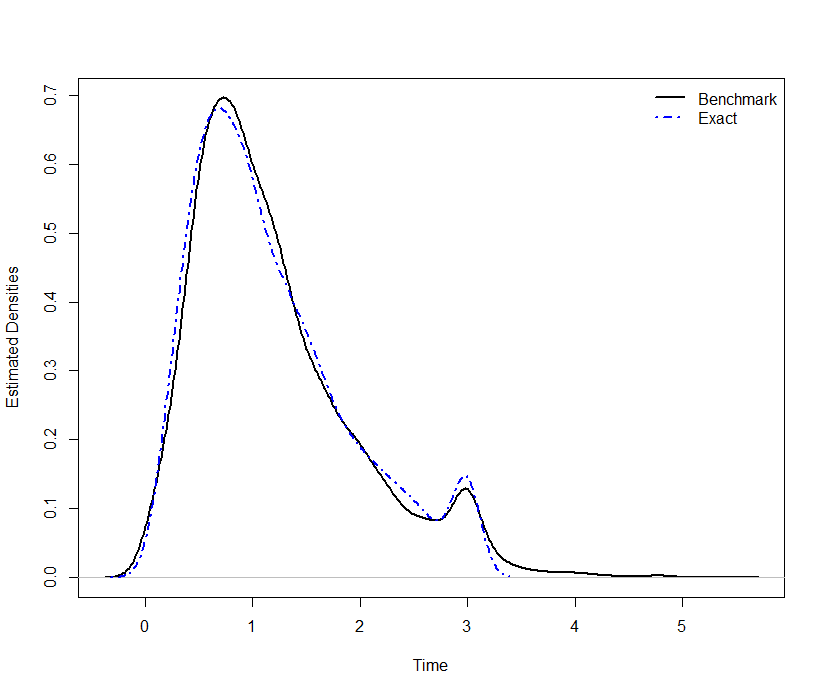}
    \caption{Estimated FPT density from $10^3$ samples of FPT. 
    Benchmark results from \cite{herrmann2023exact} are shown in solid black, while the HEx Scheme with $\epsilon=10^{-3}$ and $s_{\min}=-1$ is shown in dashed blue.}
    \label{fig:benchmark_example}
\end{figure}

\noindent Equivalently, the Markov kernel $K:E\times\mathcal B(E)\to[0,1]$ reads, for $A\in\mathcal B(\mathbb{R})$,

\begin{equation*}
K\!\big((y,n),\,A\times\{n+1\}\big)
\;=\;
\int_{\mathcal E}
\mathbf 1_A\!\big(y + j(t,y,\eta)\big)\,\nu(d\eta),
\qquad
K\!\big((y,n),\,\mathbb{R}\times\{n\}\big)=0 .    
\end{equation*}
Figure \ref{fig:benchmark_example} shows the estimated FPT densities obtained from $10^3$ FPT samples using both methods. The two density curves exhibit a close behaviour. The minor discrepancies visible between the curves lie within statistical variation. This is confirmed by the Kolmogorov-Smirnov (KS) test (p value = 0.3877). Furthermore, in this example we observe the minimum slope $s_{\text{min}}=-1$ is enough to pass the KS-test. This indicates that, in many practical settings, suitable parameters can be identified that enable the HEx Scheme to produce results nearly indistinguishable from the true values.

\section{Neuroscience application: predicting spike times in a neuron using the HEx scheme}
\label{sec:application}
Spikes constitute the fundamental mechanism by which neurons transmit and encode information. Their generation is governed by the membrane potential, which represents the difference in electric potential between the inside and outside of the cell membrane. This potential fluctuates due to the flow of ions through various channels, influenced by synaptic input and intrinsic cellular dynamics. When the potential reaches a certain threshold, a rapid depolarisation occurs, producing a spike, after which the potential resets toward its resting value before evolving again. If the threshold is not reached, the membrane potential continues to fluctuate without producing a spike and eventually settles back to its resting level. For more details, we refer the reader to \cite{ermentrout2010mathematical}.\\
\noindent The most classical mathematical description of this mechanism is the leaky integrate-and-fire (LIF) model, a linear SDE with a constant threshold cf. \cite{ditlevsen2007parameters, buonocore2010stochastic}. Here, we adopt a more realistic variant of the model called the quadratic leaky integrate-and-fire (QLIF), \cite{khurana2024exact, gerstner2014neuronal}, which captures the non-linear restorative dynamics of the membrane potential between spikes. \\
Neurons receive numerous input signals at synapses, with their influence depending on their proximity to the cell body. Distal synaptic inputs, being numerous and individually weak, can be well approximated by a diffusion term in the QLIF equation. However, inputs from proximal synapses, in particular those close to the cell body, are few but have large amplitudes, making the diffusion approximation not feasible. To capture their impact, we follow the approach in \cite{lansky2008review, giraudo2002effects, giraudo1997jump}, introducing discrete jump terms to represent excitatory and inhibitory synaptic inputs on top of the continuous diffusion dynamics. The resulting system takes the form of a JD model for the membrane potential.\\
Let $V_t$ denote the membrane potential, governed by the JD
\begin{equation}
    dV_t = \left(-\frac{1}{\mathcal{T}}V_t(V_t-V_\mathrm{rest}) + I V_t\right)dt 
    + \sigma V_t\,dB_t 
    + aV_t\,dP_t^{+} - bV_t\,dP_t^{-}, 
    \qquad V_0 = v_0,
    \label{memb_pot}
\end{equation}
where $V_\mathrm{rest}$ is the resting potential, $I$ the input current, and $B_t$ a standard Brownian motion. 
The independent Poisson processes $P^{+}$ and $P^{-}$, with respective intensities $\lambda^{+}$ and $\lambda^{-}$, represent excitatory and inhibitory synaptic events. 
The drift combines a leak term $-\frac{1}{\mathcal{T}}V_t(V_t-V_\mathrm{rest})$, which drives the potential back to resting potential, and an input term $I V_t$ that pushes it toward the firing threshold.\\
Let $(\Omega, \mathcal{F}, (\mathcal{F}_t)_{t\ge0}, \mathbb{P})$ be a filtered probability space and let $B_t$ be a standard Brownian motion and $p_\phi(d\eta \times dt)$ be a Poisson random measure with compensator $\phi(d\eta)\,dt$. We denote the mark space by $\mathcal{E}=\{+, -\}$, where the $"+"$ denotes an excitatory input and $"-"$ an inhibitory one and 
and
\begin{equation*}
    \phi(d\eta)=\lambda^{+}\,\delta_{+}(d\eta)+\lambda^{-}\,\delta_{-}(d\eta), 
\qquad \lambda^{\pm}>0.
\end{equation*}
\noindent The process state is $U_t=(V_t, N_t)\in E=\mathbb{R}\times\mathbb{N}$, where $V_t$ is the membrane potential and $N_t$ counts synaptic events. Between successive jumps $\{T_i\}_{i\ge1}$, $V_t$ satisfies the continuous SDE
\begin{equation*}
    dV_t = \mu(V_t)\,dt + \sigma(V_t)\,dB_t,
\quad \text{with}\quad 
\mu(v)=-\frac{1}{\mathcal{T}}v(v-V_\mathrm{rest}) + I v, 
\quad \sigma(v)=\sigma v.
\end{equation*}
\noindent Jumps occur at rate $\lambda=\lambda^{+}+\lambda^{-}$.  
At each jump time $T_i$, an independent mark $\eta_i$ is drawn from the normalized measure $\nu(d\eta)=\phi(d\eta)/\lambda$.  
The jump map $j:\mathbb{R}\times\mathcal{E}\to\mathbb{R}$ acts as
\begin{equation*}
 V_{T_i}=j(V_{T_i^{-}},\eta_i)
=\begin{cases}
(1+a)V_{T_i^{-}}, & \eta_i=+,\\[2pt]
(1-b)V_{T_i^{-}}, & \eta_i=-,
\end{cases}
\qquad N_{T_i}=N_{T_i^{-}}+1.   
\end{equation*}
\noindent This mechanism is equivalently described by the transition kernel
\begin{equation*}
  K\big((v,n),A\times\{n+1\}\big)
=\tfrac{\lambda^{+}}{\lambda}\mathbf{1}_A\big((1+a)v\big)
+\tfrac{\lambda^{-}}{\lambda}\mathbf{1}_A\big((1-b)v\big),
\qquad A\in\mathcal{B}(\mathbb{R}).  
\end{equation*}
To reflect adaptation after each spike, we employ an exponentially decaying threshold as in \cite{hultborn1979recurrent, levakova2019adaptive}
\begin{equation*}
\theta(t) = \theta_0 + e^{-t}, \quad \theta_0>v_0.    
\end{equation*}
\noindent Our aim is to investigate the effect of input contributions arising from synapses located near the cell body. To this end, we simulate spike times using the HEx Scheme for different combinations of model parameters $a, b , \lambda^{+}$, and $\lambda^{-}$. The influence of the input current $I$ in the drift term has been examined previously in \cite{khurana2024exact}.
For each parameter set, we compute and plot the firing rate, defined as $1/\mathbb{E}[\tau_b]$, where $\tau_\theta = \inf\{t \geq 0: V_t \geq \theta(t)\}$.\\
\noindent Since the JD includes two independent Poisson processes, one for excitatory and one for inhibitory events, the HEx Scheme (in Algorithm \ref{alg:HEx Scheme}) is then adapted to handle multiple jump sources. \\
Let $t_1^{+}$ and $t_1^{-}$ denote the first jump times of the excitatory and inhibitory processes, respectively, and define
$T_1 = \min(t_1^{+}, t_1^{-})$ as the time of the first event.  
At each subsequent iteration, only the jump time associated with the process that generated the last event is updated, while the other is kept unchanged. For instance, if $t_i^{+} = T_i$, then a new $t_{i+1}^{+} = T_i + e^{+}$ with $e^{+} \sim \mathrm{Exp}(\lambda^{+})$ is generated, while $t_{i+1}^{-} = t_i^{-}$ is retained, the converse applies if $t_i^{-} = T_i$. The next event time is then given by
\begin{equation*}
  T_{i+1} = \min(t_{i+1}^{+}, t_{i+1}^{-}),  
\end{equation*}
and the procedure continues recursively.  

\noindent The associated tracking process is given by
\begin{equation}
   dV_t^{i,\infty} = \left(-\frac{1}{\mathcal{T}}V_t^{i,\infty}(V_t^{i,\infty}-V_\text{rest}) + I\cdot V_t^{i,\infty}\right)dt+\sigma V_t^{i,\infty}dB_t, \quad t\in [T_i,\infty).
   \label{eq: tracking process application}
\end{equation}
To obtain a unit-diffusion SDE, we apply the transformation from \cite{beskos2006retrospective}:
\begin{equation}
X_t^{i,\infty} = -\frac{1}{\sigma} \log (V_t^{i,\infty})
\label{eq:app_transformation}
\end{equation}
which yields
\begin{equation}
    dX_t^{i,\infty}=\left(\frac{\sigma}{2}-\frac{V_\mathrm{rest}}{\tau \sigma}-\frac{I}{\sigma} + \frac{1}{\mathcal{T} \sigma}e^{-\sigma X_t^{i,\infty}}\right)dt + dB_t, \quad t\in [T_i,\infty),
\end{equation}
and the corresponding threshold
\[
 \beta(t)=-\frac{1}{\sigma}\log (\theta(t)).
\]
Since the threshold $\theta$ is initially above the process $V_t$, the transformed process $X_t^{i,\infty}$ remains below its threshold $\beta(t)$, requiring the following updated Bessel bridge
\begin{equation*}
  R_h = -\beta(t_c - h) + W_{t_c - h}, \qquad 0 \leq h \leq t_c - T_i. 
\end{equation*} 
Next, we verify that the transformations applied at the jump times (Steps 8 and 10 of Algorithm~\ref{alg:HEx Scheme}) are well defined. At each jump time (Step 8), the inverse transformation yields
\begin{equation*}
V_{T_{i+1}}^{i,\infty}=e^{-\sigma X_{T_{i+1}}^{i,\infty}}, \qquad 
V_{T_{i+1}}
=\begin{cases}
(1+a)e^{-\sigma X_{T_{i+1}}^{i,\infty}}, & T_{i+1}\text{ from }P^+,\\[3pt]
(1-b)e^{-\sigma X_{T_{i+1}}^{i,\infty}}, & T_{i+1}\text{ from }P^-,
\end{cases}    
\end{equation*}

and re-applying the transformation (Step 10) gives
\begin{equation*}
X_{T_{i+1}}^{i+1,\infty}
=X_{T_{i+1}}^{i,\infty}+\tfrac{1}{\sigma}\log(1+\eta_i),    
\end{equation*}

which is well defined for $0\le b<1$.

\begin{figure}[H]
  \centering
  \begin{subfigure}[b]{0.48\textwidth}
    \centering
    \includegraphics[width=\linewidth]{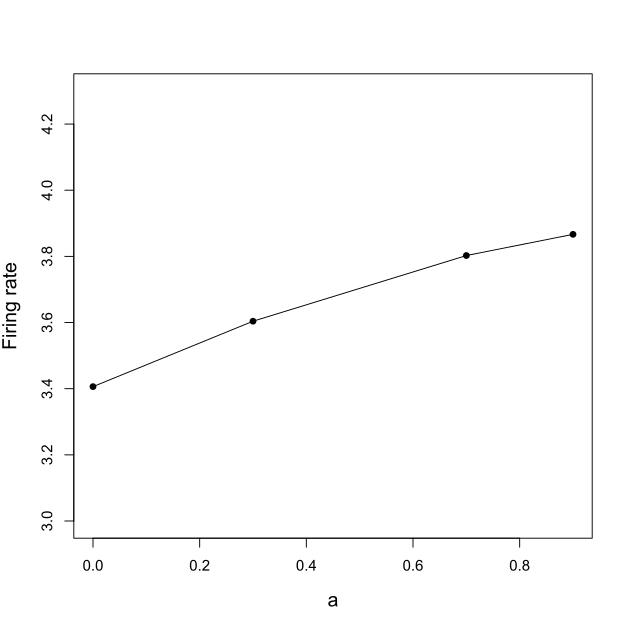} 
    \caption{Varying the positive jump size $a$ while keeping other parameters constant: $b=0.5$, $\lambda^{+}=1$, $\lambda^{-}=1$.}
    \label{fig:plot1}
  \end{subfigure}\hfill
  \begin{subfigure}[b]{0.48\textwidth}
    \centering
    \includegraphics[width=\linewidth]{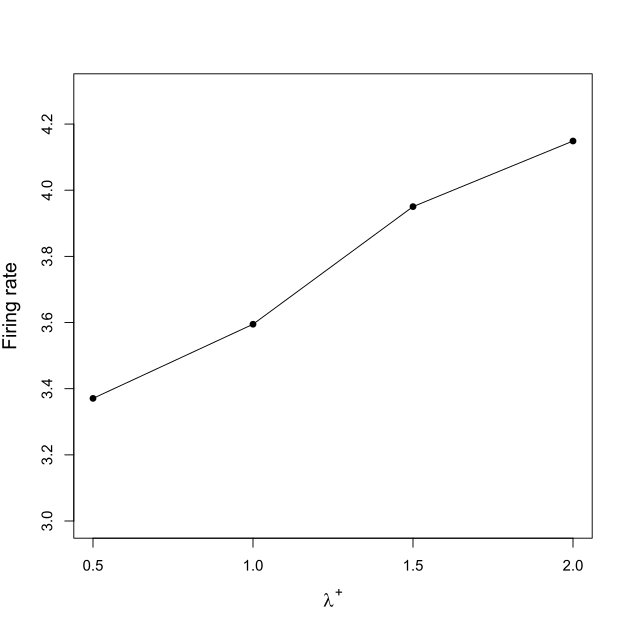} 
    \caption{Varying the jump rate $\lambda^{+}$ associated with positive jumps, while keeping the other parameters constant: $a=0.5$, $b=0.5$, $\lambda^{-}=1$.}
    \label{fig:plot2}
  \end{subfigure}
  \caption{Effect of positive jump parameters on the firing rate. The firing rate of a neuron is estimated from $2\cdot 10^3$ spike time samples using HEx Scheme. The parameters for the continuous part of the process were taken as follows: $\tau=2$, $\sigma=1$, $V_\mathrm{rest}=1$, $I=3$, $v_0=1/e$.}
  \label{fig:application_plots_1}
\end{figure}

\begin{figure}[htb]
  \centering
  \begin{subfigure}[b]{0.48\textwidth}
    \centering
    \includegraphics[width=\linewidth]{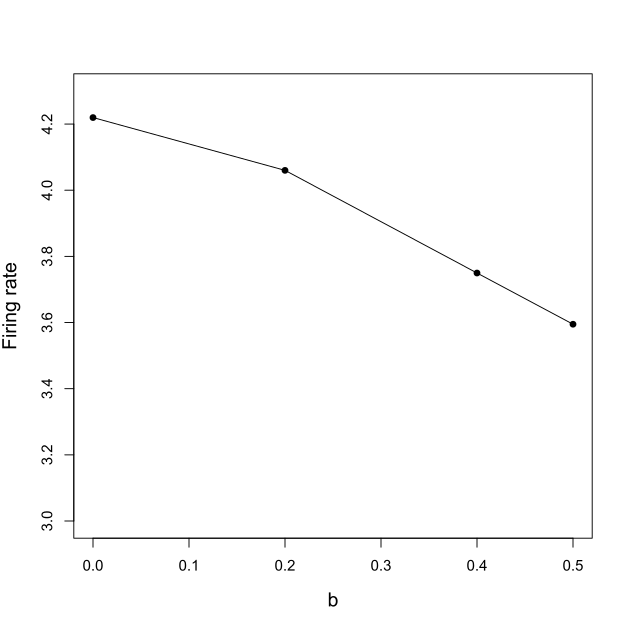} 
    \caption{Varying the negative jump size $b$ while keeping other parameters constant: $a=0.5$, $\lambda^{+}=1$, $\lambda^{-}=1$.}
    \label{fig:plot3}
  \end{subfigure}\hfill
  \begin{subfigure}[b]{0.48\textwidth}
    \centering
    \includegraphics[width=\linewidth]{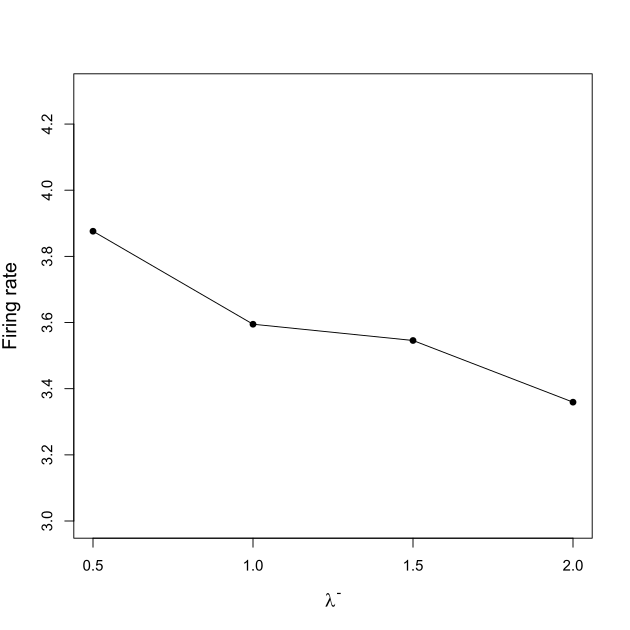} 
    \caption{Varying the jump rate $\lambda^{-}$ associated with negative jumps, while keeping other parameters constant: $a=0.5$, $b=0.5$, $\lambda^{+}=1$.}
    \label{fig:plot4}
  \end{subfigure}
  \caption{Effect of negative jump parameters on the firing rate. The firing rate of a neuron is estimated from $2\cdot 10^3$ spike time samples using HEx Scheme. The parameters for the continuous part of the process were taken same as in Figure \ref{fig:application_plots_1}.}
  \label{fig:application_plots_2}
\end{figure}

\noindent Figures \ref{fig:application_plots_1} and \ref{fig:application_plots_2} illustrate the effect of adding jumps to the neuron model. We simulate spike time samples and evaluate the firing rate for different values of the parameters that define the jumps. We observe in Figure \ref{fig:plot1} that larger positive jump sizes increase the firing rate, as these push the membrane potential closer to the threshold. Biologically, this corresponds to stronger excitatory inputs leading to more frequent spikes. Conversely, larger negative jump sizes produce the opposite effect, reducing the firing rate, see Figure \ref{fig:plot3}.\\
We also observe in Figure \ref{fig:plot2} that as the positive jump rate $\lambda^{+}$ increases, the firing rate also increases. This is expected since a higher number of positive jumps pushes the membrane potential more often toward the threshold. Biologically, this indicates that when more excitatory inputs from synapses close to the cell body occur, spikes become more frequent. Conversely, more negative jumps produce the opposite effect, reducing the firing rate, see Figure \ref{fig:plot4}.
\begin{remark}
    \begin{itemize}
        \item[(i)] It is noteworthy that the Exact simulation method works in case of a quadratic drift term, whereas time-discretization methods often lead to numerical instabilities (cf.~\cite{szpruch2010numerical}). Even small errors in the drift term may accumulate over time and cause blow-ups.
        \item[(ii)] The HEx Scheme further enables the treatment of more general and realistic cases involving time-dependent thresholds, such as the exponentially decaying threshold considered here.
    \end{itemize}
\end{remark}

\section{Conclusion}
In this work, we established a correspondence between the JD model and PDifMPs. This connection allowed us to apply the method developed for the FPT of PDifMPs with time-dependent thresholds to JD models. The PDifMP representation is particularly convenient for numerical visualization and analysis, as it naturally separates the process into three components: the continuous part, the jump rate, and the transition kernel defining the jump mechanism. We explicitly identified these components for a JD.\\
We compared the proposed algorithm with the constant-threshold version to examine the modifications required for time-dependent boundaries. The analysis revealed additional computational steps, for which we derived bounds on the expected number of extra iterations.\\
We also applied the proposed algorithm to a neuron model whose continuous part follows an SDE with quadratic drift and includes two independent jumps. Time-discretization methods are known to be less reliable for models with higher-order coefficients, highlighting that our method would be efficient to use in such models.\\
Future work in this direction could be to extend the method for processes with stochastic resetting and scenarios with two-sided, time-dependent thresholds.

\newpage
\bibliographystyle{plain}
\bibliography{Ref}
\end{document}